\documentstyle[psfig]{l-aa}

\begin{document}

\thesaurus{03(04.19.1;12.04.2;13.25.2)}

   \title{The ROSAT Deep Survey}

   \subtitle{I. X--ray sources in the Lockman Field}
 
   \author{G. Hasinger\inst{1}, 
           R. Burg\inst{2}
           R. Giacconi\inst{3}
           M. Schmidt\inst{4}
           J. Tr\"umper\inst{5}
           G. Zamorani\inst{6,7}}
 
   \offprints{G. Hasinger}

\institute{Astrophysikalisches Institut Potsdam, An der Sternwarte 16,
     14482 Potsdam, Germany
\and Johns Hopkins University, Baltimore, MD 21218, USA
\and European Southern Observatory, Karl--Schwarzschild--Str. 1,
     85748 Garching bei M\"unchen, Germany
\and California Institute of Technology, Pasadena, CA 91125, USA
\and Max--Planck--Institut f\"ur extraterrestrische Physik,
     Karl--Schwarzschild--Str. 2, 85740 Garching bei M\"unchen, Germany
\and Istituto di Radioastronomia del CNR, via Gobetti 101,
     I--40129 Bologna, Italy
\and Osservatorio Astronomico, via Zamboni 33, I-40126 Bologna, Italy}

\date{Received 12 May 1997; accepted 4 Aug 1997}
 
   \maketitle


   \begin{abstract}
 
The ROSAT Deep Survey in the {\it Lockman Hole} is the most sensitive X--ray 
survey performed to date, encompassing an exposure time of 207 ksec with 
the PSPC and a total of 1.32 Msec with the HRI aboard ROSAT. Here we 
present the complete catalogue of 50 X-ray sources with PSPC fluxes (0.5--2 keV) 
above $ 5.5 \times 10^{-15}~erg~cm^{-2}~s^{-1}$. The optical identifications 
are discussed in an accompanying paper (Schmidt et al., 1997). We also derive 
a new log(N)--log(S) function reaching a source density of $970 \pm 150 ~deg^{-2}$ 
at a limiting flux of $10^{-15}~erg~cm^{-2}~s^{-1}$. At this level 70-80\% 
of the 0.5--2 keV X-ray background is resolved into discrete sources. Utilizing 
extensive simulations of artificial PSPC and HRI fields we discuss in detail
the effects of source confusion and incompleteness both on source counts and 
on optical identifications. Based on these simulations we set 
conservative limits on flux and on off-axis angles, which guarantee a high
reliability of the catalogue. We also present simulations of shallower 
fields and show that surveys, which are based on PSPC exposures longer than 
50 ksec, become
severely confusion limited typically a factor of 2 above their $4\sigma$ 
detection threshold. This has consequences for 
recent claims of a possible new source population emerging at the 
faintest X-ray fluxes.

\keywords{surveys -- cosmology: diffuse radiations -- X-rays: galaxies}

\end{abstract}
 
\section{Introduction}

The study of the X--ray background (XRB), originally discovered 35 years ago
(Giacconi et al., 1962), with imaging X--ray telescopes has progressed 
rapidly  
in the last few years. After a pioneering start with deep Einstein exposures 
(Giacconi et al., 1979, Griffiths et al., 1983, Hamilton et al., 1991), 
ROSAT deep survey
observations were able to resolve the majority of the soft XRB into discrete
sources (Hasinger et al., 1993, hereafter H93). First deep imaging exposures with ASCA have
extended these studies into the hard X--ray band (Inoue et al. 1996,
Cagnoni et al., 1997). 
The X--ray band is one of the 
few regions of the electromagnetic spectrum, where the integrated emission of
discrete sources, summed over all cosmic epochs, is dominating the celestial
emission. The XRB therefore provides a strong constraint for cosmological 
evolution
models (see e.g. Comastri et al., 1995; Zdziarski 1996).
In order to understand the populations 
contributing to the X--ray background and their cosmological evolution
it is necessary to obtain complete optical identification of the sources
detected in X--ray deep surveys.
The majority of optically identified X--ray sources at faint fluxes
are active galactic nuclei (AGN), i.e. QSOs and Seyfert galaxies with 
broad emission lines (Shanks et al., 1991; Boyle et al., 1993; Bower et 
al., 1996; Page et al., 1996). Using the large samples of X--ray selected AGN
now becoming available, their luminosity function and its evolution with 
redshift can be studied in detail (Boyle et al., 1995; Page et al., 1996; 
Jones et 
al., 1996). The current models, which are based on pure luminosity evolution,
predict that the broad--line AGN contribute only 30--50\% to the soft X--ray
background and much less to the hard X--ray background. Postulating a large
population of faint, intrinsically absorbed AGN, 
some models predict a much larger fraction of the soft
XRB, and possibly all of the hard XRB as due to AGN (see e.g. Comastri et al.,
1995; Zdziarski 1996).  
 
There are, however, reports that at the faintest X--ray fluxes a new population
of X--ray active, but optically relatively normal narrow--emission line 
galaxies (NELG) start to dominate the X--ray counts, and could 
ultimately contribute the majority of the XRB (Georgantopoulos
et al. 1996, Griffiths et al. 1996, McHardy et al., 1997). These claims are 
based on deep ROSAT
PSPC surveys which -- as we will show -- start to be severely confusion limited
at the faintest fluxes. Care has therefore to be taken to assess
the systematic position errors and misidentifiction likelihood at faint
fluxes. In our own deep survey
identification work we have therefore taken measures to minimize 
confusion and to increase the reliability of the X--ray catalogue. 
We restricted the sample to X--ray fluxes well above the detection threshold
at a reliable flux level derived from detailed simulations. 
Secondly, we invested massive amounts of
ROSAT HRI time to cover our deepest survey field in the Lockman Hole, this
way minimizing position errors and confusion. In this paper (paper I) we
describe the derivation of the X--ray source catalogue in the Lockman
Hole. Section 2 gives a summary of the observations, Sect. 3 describes
the new maximum likelihood (ML) crowded--field detection algorithm developed
for deep X-ray 
survey work, Sect. 4 discusses the necessary astrometric corrections for
ROSAT PSPC and HRI images and Sect. 5 gives the final source catalogue.
The verification of the analysis procedure through simulations and the
derivation of a final log(N)--log(S) function are discussed in Sect. 6
and a discussion of the results is given in Sect. 7. The complete optical 
identification of the X--ray source catalogue is described in {\it paper II} 
(Schmidt et al., 1997) and deep VLA radio observations of the field
are discussed by deRuiter et al. (1997).

\section{Observations}

\begin{figure}[htp]
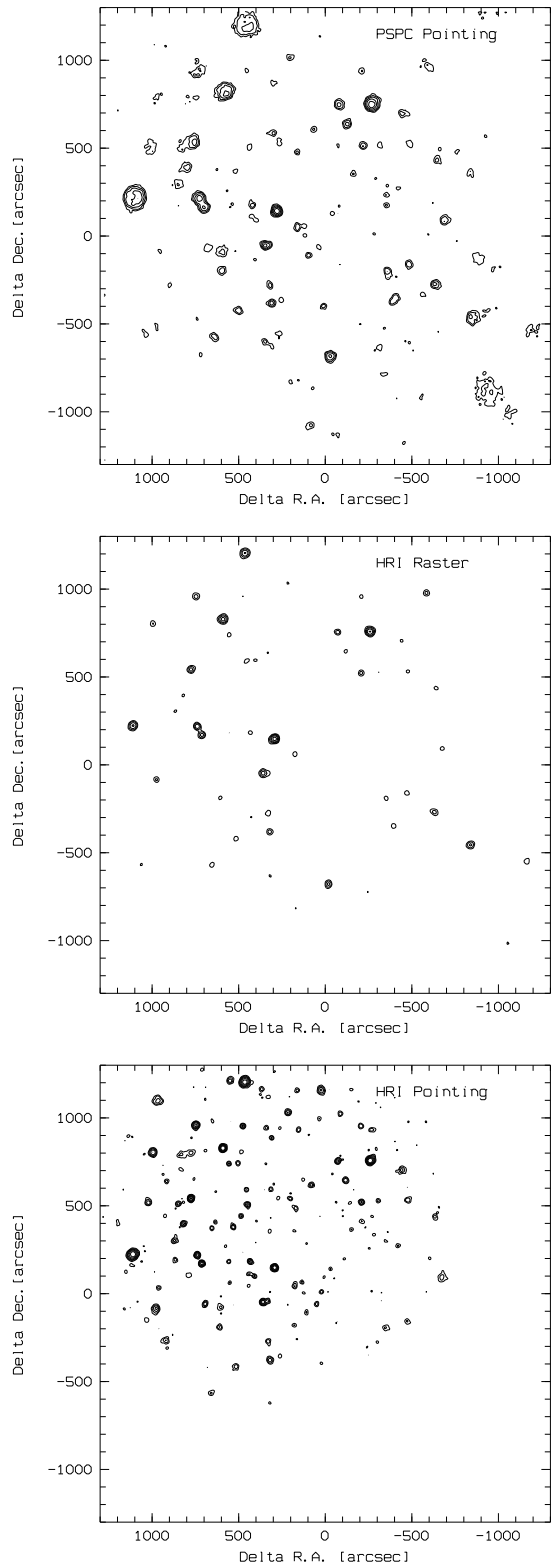

\vskip -1.0truecm
\unitlength1cm
\begin{minipage}[t]{8.0cm}
\begin{picture}(8.0,8.0)
\psfig{figure=h0499.f1,width=8.0cm,angle=-90}
\end{picture}\par
\end{minipage}
\hfill
\vskip -1.0truecm
\begin{minipage}[t]{8.0cm}
\begin{picture}(8.0,8.0)
\psfig{figure=h0499.f2,width=8.0cm,angle=-90}
\end{picture}\par
\end{minipage}
\hfill
\vskip -1.0truecm
\begin{minipage}[t]{8.0cm}
\begin{picture}(8.0,8.0)
\psfig{figure=h0499.f3,width=8.0cm,angle=-90}
\end{picture}\par
\vskip -0.2 truecm
\end{minipage}
\caption[ ]{\small ROSAT Deep Survey images in the Lockman Hole: 
(a) PSPC 207 ksec pointing, (b) HRI 205 ksec raster scan and 
(c) HRI 1112 ksec pointing. The images are all centered at the 
PSPC pointing direction (see text).}
\label{IMA}
\end{figure}

The data presented here consists of all observations of the ROSAT 
Deep Survey accumulated in the period 1990--1997 in the direction of the 
Lockman Hole, which is one of the areas on the sky with a minimum of
the galactic Hydrogen column density; $N_H$ is roughly $5 \times
10^{19}~cm^{-2}$ in this field (Lockman et al., 1986). Table 1 gives a summary 
of these observations. Specifically, a total exposure time of 207.41 ksec 
was accumulated with PSPC (Pfeffermann and Briel 1992) pointings centered
at the direction $RA(2000)=10^h52^m$, $DEC(2000)=57^o21'36"$. These data have 
been partially 
presented in H93. For the PSPC data all good time intervals
selected by the ROSAT standard analysis (SASS; Voges 1992) have been 
analysed. The individual datasets obtained in different semesters, 
which are affected by residual erratic aspect errors of order 10 arcsec, 
were all shifted to a common celestial reference system using preliminary
optical identifications. A PSPC image with a pixel size of 5
arcsec, covering the most sensitive area of the field of 
view (FOV), with a radius of $\sim 20$ arcmin, has been accumulated in the 
standard PSPC ``hard'' energy band 0.5--2 keV (pulseheight channel 52--201).
This choice optimizes the angular resolution of the PSPC and the signal to
noise ratio for the detection of faint extragalactic sources. It is
also relatively insensitive to neutral hydrogen absorption column
densities below $\sim 10^{21}~cm^{-2}$. 
For the calculation of hardness ratios we additionally analysed the PSPC 
data in the ``soft'' band and further subdivided the hard band into two: H1 
and H2. Details about the energy bands are given in Table 2. Fig. \ref{IMA}a 
shows a contour plot of the PSPC image in the hard band. 

The same area has also been covered by a set of raster 
scan observations with the ROSAT HRI (David et al., 1996). A total of $\sim
100$ pointings of roughly 2 ksec exposure each, has been placed on a regular
grid with a step size of $\sim 2$ arcmin. The HRI FOV has a size
of roughly $36 \times 36$ arcmin. The inner, most sensitive part of the field,
where the off--axis blur of the telescope can be neglected, has a radius of
roughly 8 arcmin. The raster scan step size is much 
smaller than the HRI FOV, thus the resulting exposure covers
the survey area smoothly, with a rather homogeneous point spread function
(PSF) across the field.
The total exposure time accumulated for the 
raster scan is 205.50 ksec, comparable to the PSPC pointing. The individual
HRI pointing data provided by the ROSAT Standard Analysis System SASS (Voges 
1992) have been merged into a single mosaique, after separate 
astrometrical solutions have been applied to each dataset, in
order to correct for residual boresight and scale factor errors in the   
HRI data (see below). The HRI raster image has been accumulated in the 
restricted pulseheight channel range 1--9, which significantly reduces the 
detector background with a minimum loss ($\sim 7\%$) of cosmic X--ray photons 
(David et al., 1996). Using this channel selection, the HRI is sensitive 
to cosmic X--ray photons in the full ROSAT energy range 0.1--2.4 keV.
Fig. \ref{IMA}b shows the resulting HRI raster scan image
on the same scale as the PSPC data.

\begin{table} 
\caption[ ]{Observation Summary}
\begin{flushleft}
\begin{tabular}{lllr}
\hline
Start Date & End Date & Instrument & Exp. [s]\\
\hline
1990 Apr 16 & 1991 May 21 & PSPC       &  67989\\ 
1991 Apr 25 & 1991 May ~5 & HRI raster &  86778\\
1991 Oct 25 & 1991 Nov ~2 & PSPC       &  24327\\
1991 Oct 27 & 1991 Nov ~9 & HRI raster & 116635\\
1992 Apr 15 & 1992 Apr 24 & PSPC       &  66272\\
1992 Apr 18 & 1992 Apr 18 & HRI raster &   2082\\
1992 Nov 29 & 1992 Nov 29 & PSPC       &   2082\\
1993 Apr 26 & 1993 May ~9 & PSPC       &  46740\\
1994 Oct 21 & 1994 Nov ~7 & HRI        & 106565\\
1995 Apr 15 & 1995 May 11 & HRI        & 294483\\
1995 Oct 26 & 1995 Nov 27 & HRI        & 200358\\
1996 May ~1 & 1996 May 29 & HRI        & 319590\\
1997 Apr 15 & 1997 Apr 28 & HRI        & 191094\\
\hline
\end{tabular}
\end{flushleft}
\end{table}

\noindent
From Fig. \ref{IMA} it is obvious, that the HRI raster scan, although
having significantly higher resolution, is not nearly as sensitive as the 
PSPC observation which, however, is reaching the confusion limit (see 
below). In an attempt to reach an unconfused sensitivity limit comparable or
deeper than the PSPC exposure, at least in a substantial fraction of the PSPC 
FOV, we performed an ultradeep HRI exposure of 1 Msec in a single 
pointing direction. 
The ROSAT attitude system shows residual pointing errors on the order of 
several arcsec, which can lead to a corresponding positional shift of the 
resulting image (see below). Uncorrected aspect errors can also 
lead to a substantial ellipticity of the HRI images of point sources 
(David et al., 1996), thus reducing re\-solution and sensitivity. To some 
degree these aspect errors can be corrected for, if one or more bright 
optically identified X--ray sources are present in the image. In order to 
allow this procedure, we selected a pointing direction for the ultradeep
HRI exposure, which is inside the PSPC field of view, but shifted about
10 arcmin to the North--East of the PSPC center, this way covering
a region containing about 10 relatively bright X--ray sources known from
the PSPC and the HRI raster scan. 
The ultradeep HRI exposure is centered at the direction 
$RA(2000)=10^h52^m43^s$, $DEC(2000)=57^o28'48"$. In some 
situations the automatic ROSAT HRI 
standard analysis system is too conservative in rejecting time 
intervals with supposedly high background. In particular using the
background--optimized HRI channel range 1--9, we found an optimum
point--source detection sensitivity including all exposure intervals
with good attitude data. For the ultradeep survey we therefore 
selected events using a custom made procedure after having joined the 
SASS accepted and rejected event files. This resulted in a net observing 
time of 1112.09 ksec, a gain of $\sim$ 20\% compared to the SASS products.
The individual datasets obtained in different semesters were shifted to 
the same reference system as the PSPC data, using preliminary
optical identifications. 
The HRI image in the pulseheight channel range 1-9 is shown
in Fig. \ref{IMA}c on the same scale as the PSPC image, which makes 
the shift to the North--East quite apparent.

\section{Detection algorithm}

The images in Fig. \ref{IMA} have been analysed using an improved 
version of the interactive
analysis system EXSAS (Zimmermann et al. 1994). The first steps of the 
detection procedure, i.e. the local detect algorithm LDETECT, the bicubic 
spline fit to the background map and the map detection algorithm MDETECT
have been described in H93. The detection threshold of
these algorithms has been set at a very low likelihood value so that the
resulting list of possible source positions contains several hundred 
spurious detections. This position list has then been fed into
the ``multi--ML'' crowded field detection and parameter estimation algorithm
first described by Hasinger et al. (1994a). This method, which has been 
implemented into EXSAS, works on binned image
data and fits superpositions of several PSF profiles on
top of an external background to sections of the data (typically sub--images
of arcmin size). Best--fit positions
and normalizations are obtained by maximizing the likelihood statistic 
$\cal L$ (Cash 1979) or, correspondingly, by minimizing the quantity  

\begin{equation}
 {\cal S} = -2 ln {\cal L} = -2 \sum_{i,j}(Y_{mod}(i,j)-N(i,j)lnY_{mod}(i,j)
\end{equation}

\noindent 
$Y_{mod}(i,j)$ is the sum of all model point source contributions plus
the background value in the image pixel [i,j] and N is the measured
number of photons in pixel [i,j]. For ROSAT PSPC and HRI data the 
multi--component PSF (Hasinger et al., 1994b; David et al., 1996) is 
approximated
by a single Gaussian function with a width increasing with off--axis angle.
The significance $\Delta {\cal S}$ of any 
of the individual point source contributions is then estimated by a
likelihood ratio test between the best fit with and without the corresponding
component. If $\Delta {\cal S}$ falls below a threshold value, the corresponding
component is omitted from the next iteration and only significant components 
are maintained. 
Since $\Delta {\cal S}$ follows a $\chi^2$ distribution, the errors of the 
best--fit parameters (68\% confidence single
parameter of interest) are determined by searching the parameter space
for which $\Delta {\cal S} = {\cal S} - {\cal S}_{min} =1$. Before the next 
region of interest is fit, the current PSF contributions
are added to the background model. 

The complete source detection and parameter estimation procedure 
is too complicated to validate its results and e.g. quantify its
detection efficiency analytically. We therefore performed extensive
simulations of artificial ROSAT fields run through the same set of
algorithms to study their properties (see below). 
Based on these simulations we have chosen a likelihood 
detection threshold of $\Delta {\cal S} = 16$, corresponding to a 
$4 \sigma$ detection. 
Formally, the probability of a spurious detection with such a threshold is
$6 \times 10^{-5}$ per resolution element, which would correspond to less
than one spurious detection per field of view. We will see in Sect. 6 that 
in practice the effective detection
limit is set by source confusion. Observed fluxes in the 0.5--2 keV band
are calculated for each source
using the detected counts, the corresponding exposure time and vignetting
correction, an energy--to--flux conversion factor (ECF) and a PSF--loss 
correction
factor (PCF) determined from simulations (see below). In order to derive
the ECF we assumed a 
power law spectrum with photon index 2 and galactic absorption, folded
through the instrument response. Table 2 gives a summary of the various 
energy bands and correction factors for different detectors. 
For the PSPC hard band, because of the restricted
energy range, the ECF is almost insensitive to $N_H$ and to 
the assumed spectral shape for a wide range of absorptions and spectral
indices. On the contrary, the HRI is very sensitive to absorption and 
spectral shape because of its lack of energy resolution. Unabsorbed 
sources with steep spectra are favoured by the HRI.

\begin{table} 
\caption[ ]{Energy bands and correction factors}
\begin{center}
\begin{tabular}{cccccc}
\hline
Detector & Band & Energy & Pulseheight & ECF$^a$ & PCF$^b$ \\
         &      & [keV]  & Channels    \\     
\hline
PSPC     &  H   & 0.5-2.0 & 52-201     & 0.836   & 0.90 \\
PSPC     &  S   & 0.1-0.4 & 11-41      & 1.519   & 0.90 \\
PSPC     &  H1  & 0.5-0.9 & 52-90      & 0.349   & 0.90 \\
PSPC     &  H2  & 0.9-2.0 & 91-201     & 0.487   & 0.90 \\
HRI      &      & 0.1-2.4 & 1-9        & 0.586   & 0.92 \\
\hline
\end{tabular}
\end{center}
\begin{list}{}{}
{\small
\item[$^{\rm a}$] Energy to counts conversion factor in $cts/s$ for a source with 0.5--2 keV flux 
$10^{-11}~erg~cm^{-2}~s^{-1}$\\
\item[$^{\rm b}$] PSF loss correction factor from simulations
}
\end{list} 
\end{table}
 
For the calculation of hardness ratios, PSPC images were also analysed in the 
soft (S) and the split hard (H1, H2) bands (see Table 2), however with 
fixed positions. The standard ROSAT soft and hard hardness ratios HR1 and HR2
have been used. 

\begin{eqnarray}
           HR1 = {{H - S} \over {H + S}} ~~~~~~~~~ 
           HR2 = {{H2 - H1} \over {H2 + H1}} 
\end{eqnarray}

\section{Astrometric corrections}

The ROSAT SASS performs various corrections to the individual PSPC
and HRI photons, including detector linearisation, optical distortion,
boresight and attitude correction. The large set of X--ray sources with
reliable optical identifications in the Lockman Hole allows for the first 
time to check and crosscalibrate with high accuracy the ROSAT astrometry
with absolute celestial positions.

\begin{table*}
\caption[ ]{Astrometric solutions}
\begin{center}
\begin{tabular}{cccccccc}
\hline
Instrument & Nr. & $E_{sys}$ & $\chi^2$ & $F(Sc)$ & $\Delta(\Phi)$ & $\Delta(RA)$ &$\Delta(DEC)$ \\
 & IDs  & ["] &     &             & ['] & ["] & ["] \\  
\hline
PSPC Pointing & 47 & 3.0 & 1.28  & 1.0020 $\pm$ 0.0007 & -2.9 $\pm$ 2.0 & -0.2$\pm$0.5 & -0.1$\pm$0.5 \\
HRI  Pointing & 32 & 0.5 & 1.05  & 0.9972 $\pm$ 0.0006 & ~1.2 $\pm$ 1.4 & -0.1$\pm$0.3 & ~0.3$\pm$0.3 \\
HRI  Raster   & 31 & 0.5 & 1.07  & 1.0004 $\pm$ 0.0006 & ~1.0 $\pm$ 1.5 & -0.3$\pm$0.4 & -0.3$\pm$0.4 \\
\\
Combined      & 43 & 0.0 & 1.36  & 1.0001 $\pm$ 0.0003 & -1.1 $\pm$ 0.7 & -0.1$\pm$0.2 & ~0.0$\pm$0.2 \\ 
\hline
\end{tabular}

\end{center}
\end{table*}

A complete sample of X--ray sources in the Lockman Hole with fluxes
down to $5.5 \times 10^{-15}~erg~cm^{-2}~s^{-1}$ has been optically
identified (Schmidt et al., 1997, paper II). The positions of the optical 
counterparts have been derived from Palomar 200" CCD drift scans
and have an absolute accuracy of $\sim 0.5$ arcsec. Using the set of 
point--like X--ray sources optically identified with AGN and stars, we 
could derive astrometrical solutions for the PSPC and HRI data, respectively.
For the deep PSPC and HRI pointing observations astrometrical solutions
were fit to the deviation between X--ray and optical positions. The 
transformations contained the following parameters: $\Delta(RA)$ and 
$\Delta(DEC)$ position shifts, $\Delta(\Phi)$ roll angle and
$F(Sc)$ scale factor correction. The best--fit parameters were 
determined by a $\chi^2$--fit, using the individual
statistical position errors of the X--ray sources and assuming an error
of 0.5" for the optical positions. In order to take account of possible
additional deviations in the X--ray positions, which may result from 
residuals in the detector linearization or from 
X--ray source confusion, we added in quadrature  a constant systematic 
position error $E_{sys}$ to the statistical error for each source. The 
size of this additional error (see column 3 of Table 3) was chosen such 
that the reduced $\chi^2$ approached a value of 1. 

Table 3 shows the result of the astrometric solutions. As expected,
the residual position shifts are all negligible, because all data 
have been shifted to a common system before. The roll angle correction
applied by the SASS is already very good; the roll angle corrections 
$\Delta(\Phi)$ in our deep datasets are all smaller than a few arcmin.
The only significant astrometrical correction which is necessary for
both PSPC and HRI is a scale factor correction of order 0.2--0.3\%. For the
PSPC this correction factor is marginally significant and can probably
be neglected for most datasets. For the HRI the scale factor error 
is consistent with the estimate derived from HRI data of M31 (David et al., 
1996), but our data is more accurate and provide the first significant
measurement of this quantity. The HRI scale factor error corresponds to 
a position deviation of 
about 4 arcsec across the HRI FOV and needs to be taken into account 
to achieve highest astrometrical accuracy. The residual systematic position
errors across the HRI FOV are very small (0.5"), while the PSPC has 
residual systematic errors of order 3", most likely because of source
confusion and the broader PSF (see below). 
 
For the HRI raster scan data, every individual short pointing was first
shifted to a common celestial reference system and corrected for the
above scale factor error, before all photons were merged to one dataset.
As a test we applied the same astrometrical transformation fit as to the
pointed observations. The fact that all fit parameters are consistent with 
their expected values and that the required residual systematic errors
are of the same order as for the individual HRI pointing shows that
the shift-- and merge procedure was successful. Finally, the last row in Table 
3 shows an astrometric fit to the X--ray source catalogue combined
from all three datasets (see below). As expected, all astrometric parameters
are consistent with their expected values. The reduced $\chi^2$ of the 
combined fit is acceptable without the addition of extra systematic errors. 
The distribution of positional errors for
uniquely identified sources is therefore as expected,
thus validating the calculation of the positions and errors.

\section{X--ray source catalogue}

The scale factor corrections determined in the previous section were 
applied to the three individual detection lists from PSPC pointing, 
HRI pointing and HRI raster scan observation and the corresponding systematic
position errors were added in quadrature to the statistical errors
determined in the multi--ML detection procedure. Finally, the three 
individual position lists were merged into a single source
list. For all detections, which were positionally coincident within
their 90\% error radii, weighted average positions and position errors
were calculated. These form the basis for the final X--ray source catalogue, 
presented in Table 4. This catalogue contains all objects with PSPC fluxes 
higher than $1.11\times 10^{-14}~erg~cm^{-2}~s^{-1}$
at off--axis angles smaller than 18.5 arcmin and all objects with 
fluxes larger than $0.56\times 10^{-14}~erg~cm^{-2}~s^{-1}$
inside an off--axis angle of 12.5 arcmin. 

Column (1) of Table 4 gives the official ROSAT name, column (2) an
internal source number. Columns (3) and (4) give the weighted average
coordinates of the X--ray sources for an equinox of J2000.0. Column (5) 
shows the $1\sigma$ position error of the source, including statistical
and systematic errors. A capital P, H, or R after the position error
indicates, whether the dominant weight to the X--ray position comes from the
PSPC pointing, HRI pointing or HRI raster scan, respectively. Column 
6 gives the 0.5--2 keV flux of the source in units of 
$10^{-14}~erg~cm^{-2}~s^{-1}$, derived from the ROSAT PSPC
hard band and its $1\sigma$ error. Column (7) and (8) give the soft and 
hard hardness ratios defined above.

\begin{table*}
\caption[ ]{Source Catalogue}
\begin{flushleft}
\begin{tabular}{lrllcrrr}
\hline
Source Name    & Nr. & R.A. (2000) & Dec. (2000) & Err. & $f_X$~~~~~~~     & HR1~~~~~~~         & HR2~~~~~~~       \\
               &     &             &             & ["]  &                  &                    &     \\
\hline
RX J105421.1+572545 &  28 & 10 54 21.1& 57 25 44.5& 1.0H & 19.90 $\pm$ 0.41 &  1.00 $\pm$ 0.06 & -0.27 $\pm$ 0.02 \\ 
RX J105131.1+573440 &   8 & 10 51 31.1& 57 34 40.4& 0.8H & 11.84 $\pm$ 0.33 & -0.24 $\pm$ 0.02 & -0.47 $\pm$ 0.02 \\ 
RX J105316.8+573552 &   6 & 10 53 16.8& 57 35 52.4& 0.9H &  8.82 $\pm$ 0.33 & -0.29 $\pm$ 0.02 & -0.40 $\pm$ 0.02 \\ 
RX J105239.7+572432 &  32 & 10 52 39.7& 57 24 31.7& 0.8H &  6.50 $\pm$ 0.30 & -0.55 $\pm$ 0.02 & -0.16 $\pm$ 0.03 \\ 
RX J105335.1+572542 &  29 & 10 53 35.1& 57 25 42.4& 0.8H &  5.13 $\pm$ 0.25 & -0.34 $\pm$ 0.03 & -0.36 $\pm$ 0.03 \\ 
RX J105331.8+572454 &  31 & 10 53 31.8& 57 24 53.9& 0.9H &  3.57 $\pm$ 0.18 & -0.18 $\pm$ 0.04 & -0.40 $\pm$ 0.03 \\ 
RX J105339.7+573105 &  16 & 10 53 39.7& 57 31 ~5.3& 0.9H &  3.56 $\pm$ 0.18 & -0.59 $\pm$ 0.02 & -0.43 $\pm$ 0.03 \\ 
RX J105020.2+571423 &  56 & 10 50 20.2& 57 14 22.8& 1.8R &  3.52 $\pm$ 0.17 & -0.44 $\pm$ 0.02 & -0.37 $\pm$ 0.04 \\ 
RX J105201.5+571044 &  62 & 10 52 ~1.5& 57 10 44.2& 1.6R &  3.20 $\pm$ 0.18 & -0.31 $\pm$ 0.03 & -0.50 $\pm$ 0.03 \\ 
RX J105247.9+572116 &  37 & 10 52 47.9& 57 21 16.3& 0.8H &  2.54 $\pm$ 0.13 & -0.73 $\pm$ 0.02 & -0.19 $\pm$ 0.06 \\ 
RX J105410.3+573039 &  20 & 10 54 10.3& 57 30 39.3& 1.4H &  2.43 $\pm$ 0.18 &  0.47 $\pm$ 0.11 & -0.19 $\pm$ 0.06 \\ 
RX J105154.4+573438 &   9 & 10 51 54.4& 57 34 38.0& 1.0H &  1.88 $\pm$ 0.15 & -0.23 $\pm$ 0.06 & -0.18 $\pm$ 0.06 \\ 
RX J105318.1+572042 &  41 & 10 53 18.1& 57 20 42.0& 2.4H &  1.87 $\pm$ 0.13 &  0.03 $\pm$ 0.10 & -0.46 $\pm$ 0.06 \\ 
RX J105344.9+572841 &  25 & 10 53 44.9& 57 28 40.5& 1.1H &  1.84 $\pm$ 0.15 & -0.09 $\pm$ 0.07 &  0.13 $\pm$ 0.09 \\ 
RX J105015.6+572000 &  42 & 10 50 15.6& 57 20 ~0.2& 4.7P &  1.69 $\pm$ 0.15 & -0.36 $\pm$ 0.06 & -0.14 $\pm$ 0.07 \\ 
RX J105046.1+571733 &  48 & 10 50 46.1& 57 17 32.8& 1.8R &  1.64 $\pm$ 0.17 & -0.08 $\pm$ 0.08 & -0.00 $\pm$ 0.07 \\ 
RX J105149.0+573249 &  12 & 10 51 49.0& 57 32 48.6& 1.0H &  1.57 $\pm$ 0.14 &  1.00 $\pm$ 0.78 &  0.28 $\pm$ 0.08 \\ 
RX J105324.6+571236 &  59 & 10 53 24.6& 57 12 35.7& 2.8P &  1.50 $\pm$ 0.13 &  0.27 $\pm$ 0.14 & -0.20 $\pm$ 0.07 \\ 
RX J105039.7+572335 &  35 & 10 50 39.7& 57 23 35.1& 1.8R &  1.48 $\pm$ 0.15 & -0.29 $\pm$ 0.06 & -0.19 $\pm$ 0.06 \\ 
RX J105348.7+573033 & 117 & 10 53 48.7& 57 30 33.5& 1.4H &  1.46 $\pm$ 0.18 &  1.00 $\pm$ 1.51 &  0.19 $\pm$ 0.10 \\ 
RX J105350.3+572710 &  27 & 10 53 50.3& 57 27 ~9.6& 2.0H &  1.39 $\pm$ 0.08 &  0.16 $\pm$ 0.17 &  0.18 $\pm$ 0.07 \\ 
RX J105008.2+573135 &  73 & 10 50 ~8.2& 57 31 34.7& 8.2P &  1.39 $\pm$ 0.13 &  0.78 $\pm$ 0.51 &  0.57 $\pm$ 0.44 \\ 
RX J105243.1+571544 &  52 & 10 52 43.1& 57 15 44.4& 1.2H &  1.32 $\pm$ 0.12 & -0.36 $\pm$ 0.06 &  0.00 $\pm$ 0.08 \\ 
RX J105108.4+573345 &  11 & 10 51 ~8.4& 57 33 45.4& 1.6H &  1.27 $\pm$ 0.12 &  0.18 $\pm$ 0.12 &  0.32 $\pm$ 0.11 \\ 
RX J105055.3+570652 &  67 & 10 50 55.3& 57 ~6 51.9& 8.1P &  1.24 $\pm$ 0.15 &  0.23 $\pm$ 0.16 &  0.01 $\pm$ 0.11 \\ 
RX J105020.3+572808 &  26 & 10 50 20.3& 57 28 ~7.8& 6.5P &  1.20 $\pm$ 0.11 & -0.06 $\pm$ 0.14 &  0.13 $\pm$ 0.11 \\ 
RX J105009.3+571443 &  55 & 10 50 ~9.3& 57 14 42.8& 6.6P &  1.17 $\pm$ 0.15 &  0.02 $\pm$ 0.06 &  0.13 $\pm$ 0.11 \\ 
RX J105230.3+573914 &   2 & 10 52 30.3& 57 39 13.8& 1.3H &  1.16 $\pm$ 0.14 & -0.03 $\pm$ 0.17 &  0.01 $\pm$ 0.11 \\ 
RX J105307.2+571506 &  54 & 10 53 ~7.2& 57 15 ~5.6& 1.9H &  1.12 $\pm$ 0.14 &  0.05 $\pm$ 0.10 &  0.04 $\pm$ 0.09 \\ 
RX J105319.0+571852 &  45 & 10 53 19.0& 57 18 51.9& 1.5H &  1.04 $\pm$ 0.14 &  0.02 $\pm$ 0.12 & -0.03 $\pm$ 0.08 \\ 
RX J105137.4+573044 &  19 & 10 51 37.4& 57 30 44.4& 1.0H &  0.99 $\pm$ 0.11 & -0.24 $\pm$ 0.07 & -0.18 $\pm$ 0.09 \\ 
RX J105114.5+571616 & 504 & 10 51 14.5& 57 16 16.0& 1.8R &  0.96 $\pm$ 0.12 & -0.38 $\pm$ 0.07 & -0.12 $\pm$ 0.09 \\ 
RX J105105.2+571924 &  43 & 10 51 ~5.2& 57 19 23.9& 1.9R &  0.94 $\pm$ 0.09 & -0.28 $\pm$ 0.07 & -0.03 $\pm$ 0.10 \\ 
RX J105225.1+572304 &  36 & 10 52 25.1& 57 23 ~3.7& 1.8R &  0.92 $\pm$ 0.08 &  0.05 $\pm$ 0.19 &  0.08 $\pm$ 0.11 \\ 
RX J105120.2+571849 &  46 & 10 51 20.2& 57 18 49.2& 1.9R &  0.89 $\pm$ 0.09 & -0.18 $\pm$ 0.09 & -0.13 $\pm$ 0.08 \\ 
RX J105127.0+571129 &  61 & 10 51 27.0& 57 11 29.0& 6.0P &  0.79 $\pm$ 0.09 & -0.07 $\pm$ 0.14 & -0.27 $\pm$ 0.13 \\ 
RX J105329.2+572104 &  38 & 10 53 29.2& 57 21 ~3.7& 1.3H &  0.78 $\pm$ 0.11 & -0.30 $\pm$ 0.08 & -0.53 $\pm$ 0.07 \\ 
RX J105248.4+571203 &  60 & 10 52 48.4& 57 12 ~2.7& 5.0P &  0.78 $\pm$ 0.10 & -0.00 $\pm$ 0.19 & -0.38 $\pm$ 0.09 \\ 
RX J105242.5+573159 &  14 & 10 52 42.5& 57 31 59.2& 1.3H &  0.72 $\pm$ 0.09 &  0.72 $\pm$ 0.52 &  0.44 $\pm$ 0.15 \\ 
RX J105244.4+571732 &  47 & 10 52 44.4& 57 17 31.9& 1.6H &  0.71 $\pm$ 0.09 & -0.35 $\pm$ 0.08 & -0.28 $\pm$ 0.10 \\ 
RX J105257.1+572507 &  30 & 10 52 57.1& 57 25 ~7.2& 0.9H &  0.70 $\pm$ 0.09 & -0.53 $\pm$ 0.05 & -0.47 $\pm$ 0.07 \\ 
RX J105117.0+571554 &  51 & 10 51 17.0& 57 15 54.1& 5.1P &  0.66 $\pm$ 0.11 &  0.04 $\pm$ 0.22 & -0.10 $\pm$ 0.14 \\ 
RX J105104.2+573054 &  17 & 10 51 ~4.2& 57 30 53.7& 1.6H &  0.62 $\pm$ 0.10 &  0.11 $\pm$ 0.19 & -0.14 $\pm$ 0.14 \\ 
RX J105244.7+572122 & 814 & 10 52 44.7& 57 21 22.2& 1.3H &  0.61 $\pm$ 0.16 &  0.68 $\pm$ 0.68 &  0.51 $\pm$ 0.18 \\ 
RX J105217.0+572017 &  84 & 10 52 17.0& 57 20 17.1& 2.0H &  0.60 $\pm$ 0.08 &  0.68 $\pm$ 0.71 &  0.85 $\pm$ 0.32 \\ 
RX J105259.2+573031 &  77 & 10 52 59.2& 57 30 30.8& 1.1H &  0.59 $\pm$ 0.10 & -0.14 $\pm$ 0.13 & -0.16 $\pm$ 0.14 \\ 
RX J105206.0+571529 &  53 & 10 52 ~6.0& 57 15 28.7& 4.6P &  0.58 $\pm$ 0.07 & -0.15 $\pm$ 0.12 &  0.82 $\pm$ 0.17 \\ 
RX J105237.7+573107 & 116 & 10 52 37.7& 57 31 ~6.5& 5.1P &  0.57 $\pm$ 0.10 &  1.00 $\pm$ 0.84 & -0.48 $\pm$ 0.09 \\ 
RX J105224.7+573010 &  23 & 10 52 24.7& 57 30 10.2& 1.5H &  0.56 $\pm$ 0.07 &  0.02 $\pm$ 0.20 & -0.41 $\pm$ 0.10 \\ 
RX J105237.9+571254 &  58 & 10 52 37.9& 57 12 53.5& 6.2P &  0.56 $\pm$ 0.08 &  0.07 $\pm$ 0.21 & -0.42 $\pm$ 0.09 \\ 
\hline
\end{tabular}
\end{flushleft}
\end{table*}

\begin{figure*}[htp]
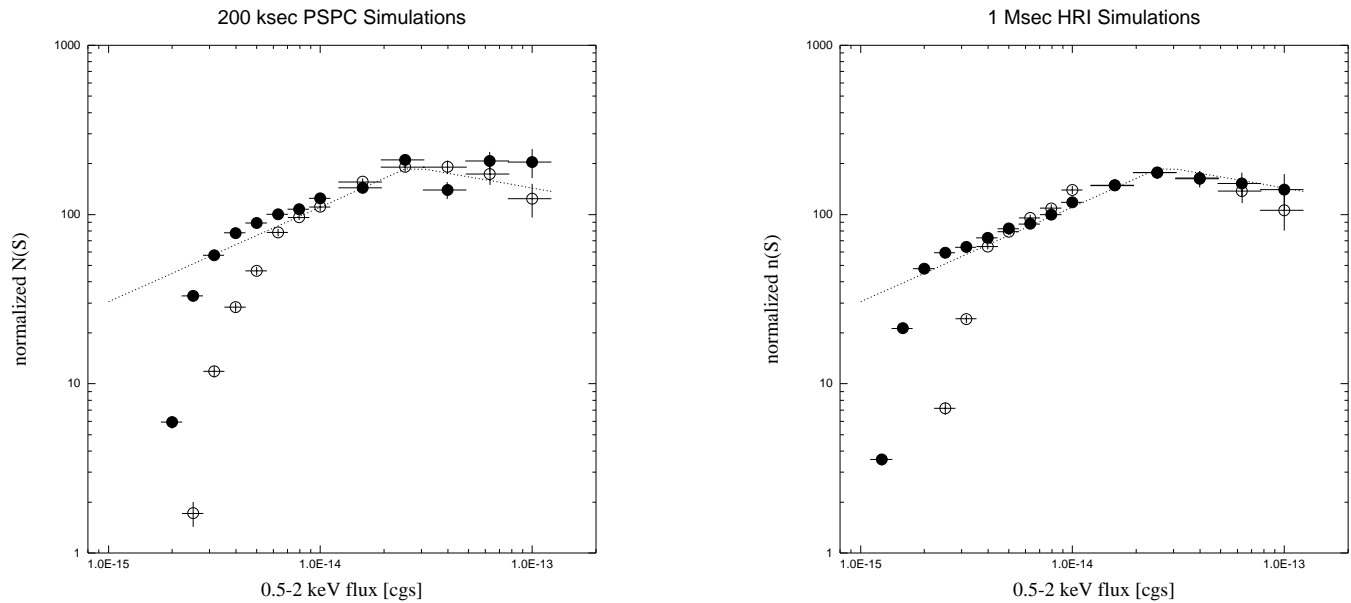

\unitlength1cm
\begin{minipage}[t]{8.0cm}
\begin{picture}(8.0,8.0)
\psfig{figure=h0499.f4,width=8.0cm}
\end{picture}\par
\end{minipage}
\hfill
\begin{minipage}[t]{8.0cm}
\begin{picture}(8.0,8.0)
\psfig{figure=h0499.f5,width=8.0cm}
\end{picture}\par
\vskip -0.2 truecm
\end{minipage}
\caption[ ]{\small Comparison of detected log(N)--log(S) relation to input
source counts in the 0.5--2 keV band. The dotted line shows the differential 
source counts n(S) input to the simulations, normalized to a Euclidean 
slope (see equation 3 and 4). The filled and open circles 
show the detected differential log(N)--log(S), derived from 
(a) 66 simulated PSPC fields of 200 ksec exposure, and (b) 100 simulated
HRI fields of 1 Msec exposure each. Filled symbols correspond to sources at 
off--axis angles smaller
than 12.5 arcmin for the PSPC and 10 arcmin for the HRI, respectively. Open
symbols refer to larger off--axis angles.}
\label{RNS}
\end{figure*}

\section{Verification of the analysis procedure by simulations}

The ROSAT deep survey exposures are probing the limits of observational and 
data analysis procedures. In order to obtain a reliable and quantitative 
characterization and calibration of the source detection
procedure, detailed simulations of large numbers of artificial fields,
analysed through exactly the same detection and parameter estimation procedure 
as the real data, are required. The simulations in H93 
demonstrated that source confusion 
sets the ultimate limit in ROSAT deep survey work with the PSPC. The 
crowded--field multi--ML detection algorithm used in the current paper was 
specifically designed to better cope with source confusion. Therefore we
felt it necessary to calibrate its efficiency and verify its 
accuracy through 
new simulations. We have simulated sets of PSPC and HRI observations with 
200 ksec and 1Msec exposure, respectively, approximating our current 
observation times. In order to compare our results with those of shallower 
PSPC surveys in the literature we simulated additional sets of observations 
of 50 and 110 ksec each. 
The soft X--ray log(N)--log(S) function (0.5--2 keV) has been simulated 
according to the ROSAT deep survey findings (H93) as a 
broken power law function for the differential source counts (all fluxes
are given in units of $10^{-14}~erg~cm^{-2}~s^{-1})$: 

\begin{eqnarray}
n(S) & = & n_1 \times S^{-b_1} \; {\rm for} \; S > S_b, \;\; 
                                 n_1 = 238.1 \;\; b_1 = 2.72 \nonumber \\ 
     & = & n_2 \times S^{-b_2} \; {\rm for} \; S < S_b, \;\;
                                  n_2 = 111.0 \;\; b_2 = 1.94 \\    
S_b  & = & 2.66          \nonumber 
\end{eqnarray}

\noindent
As shown in H93, the final results are 
relatively independent from the actual log(N)--log(S) parameters
chosen for the artificial fields.
In the simulations point sources are placed at random within the FOV, 
with fluxes drawn at random from the log(N)--log(S) function 
down to a minimum source flux of $4.65 \times 10^{-18}$, where 
formally all the X-ray background is resolved for the assumed source counts. 
(For the later comparison between input and output catalogues only input 
sources with fluxes larger than $10^{-16}$ or more than 5 simulated photons
are maintained in the input catalogue.) For each source the 
ROSAT vignetting correction, the corresponding ECF (see Table 2) and the 
exposure time (50, 110, 200 ksec for PSPC, 1Msec for HRI) are applied to the 
source flux to obtain the expected number of source counts. The actual 
source counts are drawn from a Poissonian distribution and folded through 
the point spread function. The realistic multi--component point spread function 
model is taken from Hasinger et al. (1994b) for the PSPC and from David et al.
(1996) for the HRI. 
Finally, all events missing in the field, i.e. particle background, 
non--source diffuse background and photons not simulated in the 
log(N)--log(S) function are added as a smooth distribution to the image. 

A total of 66 PSPC fields of an exposure of 200 ksec, 50 fields of
50 ksec and 27 fields of 110 ksec each has been simulated in the 
0.5--2 keV band. The corresponding 
number of HRI fields with 1 Msec exposure was 100.
All simulated exposures have been analysed through exactly the same sequence
of detection, background estimation and parameter estimation algorithms
as the real data.  
The 200 ksec PSPC images were analysed in two ranges of off--axis angles:
0--12.5 arcmin and 12.5--18.5 arcmin, the HRI 1Msec images 
for off--axis angle ranges of 0--10 arcmin and 10--15 arcmin and the 
50 and 110 ksec PSPC exposures inside 15 arcmin. 

\subsection{Flux conservation}

For each detected source the process of ``source identification'' has been
approximated by a simple positional coincidence check. A detected source was
identified with the counterpart from the input list, which appeared closest
to the X--ray position within a radius of 30 arcsec.
A direct comparison between the detected and simulated number of photons for
sources with fluxes brighter than $5 \times 10^{-14}~cgs$ shows, that 
the scale of the output flux is the same as that of the input flux to an 
accuracy of better than 1\%. This means
that at bright fluxes no systematic errors exist, apart from the PSF loss 
factor of 0.90 
(see Table 2), which is corrected for globally in the simulations as well as 
in the real data. This correction is necessary, probably due to the 
misfit between the broader components of the real PSF and the simple Gaussian 
model
in the multi-ML fit. For the HRI this PSF loss factor is 0.92 (see Table 2). 

\subsection{Limiting sensitivity and log(N)--log(S) relation}

A log(N)--log(S) function has been derived for the simulated detected sources (in rings 
of 0--12.5 arcmin and 12.5--18.5 arcmin for the PSPC and rings of 0--10 arcmin
and 10--15 arcmin, respectively for the HRI). For easier comparison the 
differential source counts n(S) were divided by the source counts expected 
for a Euclidean distribution, i.e.:

\begin{equation}
           n(S)_{cor} = n(S) \times S^{2.5}  
\end{equation}

\noindent
Fig. \ref{RNS} shows the comparison between the input log(N)--log(S) and
the detected number counts in this representation for both 200 ksec PSPC and 
the 1 Msec HRI
simulations. 
The faintest sources detected in the PSPC at small off--axis angles have a flux 
of $2 \times 10^{-15}~cgs$. For a flux lower than $3 \times 10^{-15}~cgs$, the 
source counts fall significantly below the simulated log(N)--log(S) function, 
while above this flux there is a slight excess. These effects are well known 
from ROSAT PSPC deep survey simulations and are due to a number of 
selection and confusion effects. At larger off--axis angles the sensitivity is
reduced correspondingly. The faintest HRI sources reach down to 
fluxes of $10^{-15}~cgs$ and the deviation between input and output source 
counts becomes insignificant above a flux of $1.8 \times 10^{-15}~cgs$ for 
small off--axis angles. Again, at larger angles the sensitivity is reduced.
The source confusion effects are much less pronounced in the HRI.

\begin{figure}
\psfig{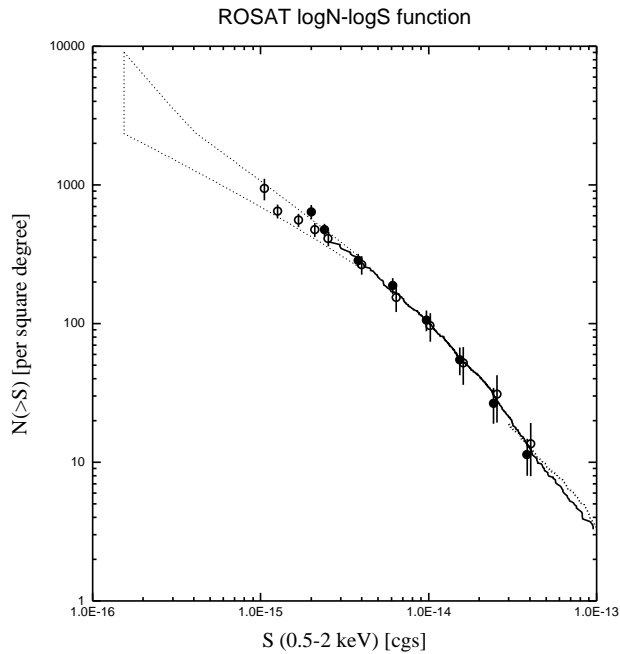}
\caption[ ]{\small Measured ROSAT log(N)--log(S) function in the Lockman Hole.
Filled circles give the source counts from the 207 ksec PSPC observation.
Open circles are from the ultradeep HRI observation (1112 ksec), slightly 
shifted in flux in order to avoid overlap of error bars. The data
are plotted on top of the source counts (solid line) and fluctuation 
limits (dotted area) from Hasinger et al. (1993). The dotted line at
bright fluxes refers to the total source counts in the RIXOS survey
(Mason et al., 1996, priv. comm.).}  
\label{LNS}
\end{figure}

The deviation between input-- and output source counts in the simulations can
be used to correct the observed log(N)--log(S) function 
down to the faintest limiting fluxes (see H93). 
Utilizing all sources detected in the real PSPC and HRI pointing observations
inside off--axis angles of 18.5 and 15 arcmin, respectively, we calculated 
the corrected observed log(N)--log(S) function displayed in Fig. \ref{LNS}.
The new PSPC data agree very well with the source counts published earlier and 
extend those down to a flux of $ 2 \times 10^{-15}~cgs$ and a surface density 
of $640 \pm 75~deg^{-2}$. The HRI log(N)--log(S) function for the first time extends 
the source counts down to fluxes of $10^{-15}~cgs$ and reaches a surface
density of $970\pm150~deg^{-2}$, about a factor of two higher than any
previous X--ray determination of source counts. All data are consistent with
the previous determination of source counts and fluctuation limits (H93).

\begin{figure*}[htp]
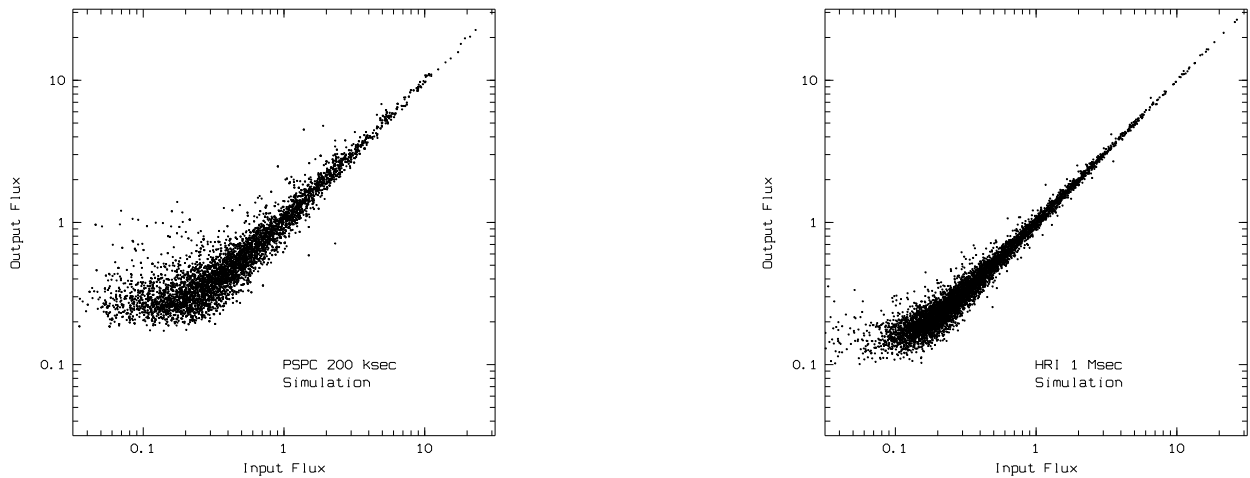

\unitlength1cm
\begin{minipage}[t]{8.0cm}
\begin{picture}(8.0,8.0)
\psfig{figure=h0499.f7,width=8.0cm,angle=-90.}
\end{picture}\par
\end{minipage}
\hfill
\begin{minipage}[t]{8.0cm}
\begin{picture}(8.0,8.0)
\psfig{figure=h0499.f8,width=8.0cm,angle=-90.}
\end{picture}\par
\vskip -0.2 truecm
\end{minipage}
\caption[ ]{\small Detected flux versus input flux for 
(a) 66 simulated PSPC fields of 200 ksec exposure, and (b) 100 simulated
HRI fields of 1 Msec exposure each. The PSPC data is for off--axis angles 
smaller than 12.5 arcmin, for the HRI the limit is 10 arcmin. Fluxes are in 
units of $10^{-14}~erg~cm^{-2}~s^{-1}$.}
\label{FLX}
\end{figure*}
 
\subsection{Source confusion}

The detected source catalogues are affected by biases and selection effects
present in the source--detection procedure. The most famous of
those is the Eddington bias, which produces a net gain of the number of
sources detected above a given flux limit as a consequence of
statistical errors in the measured flux (see discussion in H93). Another
selection effect, most important in the deep fields considered
here, is source confusion. The net effect of source confusion
is difficult to quantify analytically, because it can affect the
derived source catalogue in different ways:
\par\noindent
(1) two sub--threshold
sources could be present in the same resolution element and thus
mimic a single detected source. This leads to a net gain
in the number of sources, similar to the Eddington bias.
\par\noindent
(2) two
sources above the threshold could merge into a single brighter
source. In this case one source is lost and one is detected at a
higher flux. Whether the total flux is conserved or
not depends on the distance between the two sources and on the details
of the source detection algorithm.
\par\noindent
(3) the detection algorithm cannot discriminate close sources with very 
different brightness, which results in a net loss of fainter sources.
\par\noindent

The effects of confusion become immediately obvious in fig \ref{FLX}
where for each 
detected source the detected flux is compared to the flux of the nearest
input source within 30". While for bright sources there is an almost perfect match,
there is a significant systematic deviation for fainter X--ray sources, where
most objects appear at fluxes significantly brighter than their input 
counterparts, which is a direct indication of confusion  
because every detected source is only associated with
one input source while its flux may be contributed from several sources. 

We can quantify ``confusion'' e.g. by looking for sources, whose 
detected X--ray flux is significantly larger than their 
input flux. This means that a substantial fraction of the detected
photons originates from other, contaminating sources. We define SR 
as the ratio of the detected flux and
the input flux increased by 3 times the statistical error:

\begin{equation}
           SR = {S_{det} \over S_{inp}} \cdot {{1} \over {1 + 3 \sigma_S}}   
\end{equation}

\noindent where $\sigma_S$ is the fractional flux error. For practical 
purposes a flux ratio of $SR > 1.5$ has been chosen, above which significant
position deviations might be expected. In Table 5, we give
the {\it fraction of contaminated sources} FC with flux ratio 
SR larger than 1.5 as a function of detected flux for different surveys. 
Significant confusion (larger than 10\%) sets in below a flux of $5 \times
10^{-15}~cgs$ for the 200 ksec PSPC observation and below $10^{-14}~cgs$ 
for the 50 ksec observation. In contrast, the HRI confusion is
negligible over almost the complete range of detected fluxes.

\begin{table*}
\caption[ ]{Confusion estimates$^a$}
\begin{center}
\begin{tabular}{cccccccccccccccc}\hline

Flux range$^b$   & \multicolumn{3}{c}{200 ksec PSPC} 
             & \multicolumn{3}{c}{200 ksec PSPC} 
             & \multicolumn{3}{c}{110ksec PSPC} 
             & \multicolumn{3}{c}{ 50 ksec PSPC} 
             & \multicolumn{3}{c}{1 Msec HRI} \\
             & \multicolumn{3}{c}{0--12.5'} 
             & \multicolumn{3}{c}{12.5--18.5'} 
             & \multicolumn{3}{c}{0--15'} 
             & \multicolumn{3}{c}{0--15'} 
             & \multicolumn{3}{c}{0--10'} \\
{[$10^{-14}~cgs$]} & FC & FL & FU & FC & FL & FU & FC & FL & FU & FC & FL & FU & FC & FL & FU \\ 
\hline
2-10     & 2 & 1 & 0 & 6 & 3 & 8 & 3 & 2 & 2 & 1 & 1 & 1 & 0 & 0 &0 \\
1-2      & 3 & 4 & 3 & 9 &14 &11 & 7 & 3 & 5 & 4 & 7 & 7 & 0 & 0 &0 \\
0.5-1    & 7 &10 & 5 &14 &38 &26 & 9 &19 &11 &11 &33 &19 & 1 & 0 &0 \\
0.2-0.5  &12 &38 &16 &16 &78 &34 &11 &62 &22 &14 &74 &19 & 2 & 3 &1 \\
0.1-0.2  &20 &81 &24 &   &   &   &   &   &   &   &   &   &11 &47 &10\\
\hline
\end{tabular}
\end{center}
$^a$ Fraction of confused sources in percent according to different 
confusion measures (see text): FC is the percentage of contaminated sources,
having an output flux significantly larger than their input flux ($SR>1.5$). 
FL is the percentage of lost sources, having no detected source within
15" from the input position. FU is the fraction of unidentifiable 
detected sources, having no input source within 15" from the detected 
position.
$^b$ detected flux for FC and FU, input flux for FL 
\end{table*}

\begin{figure*}[htp]
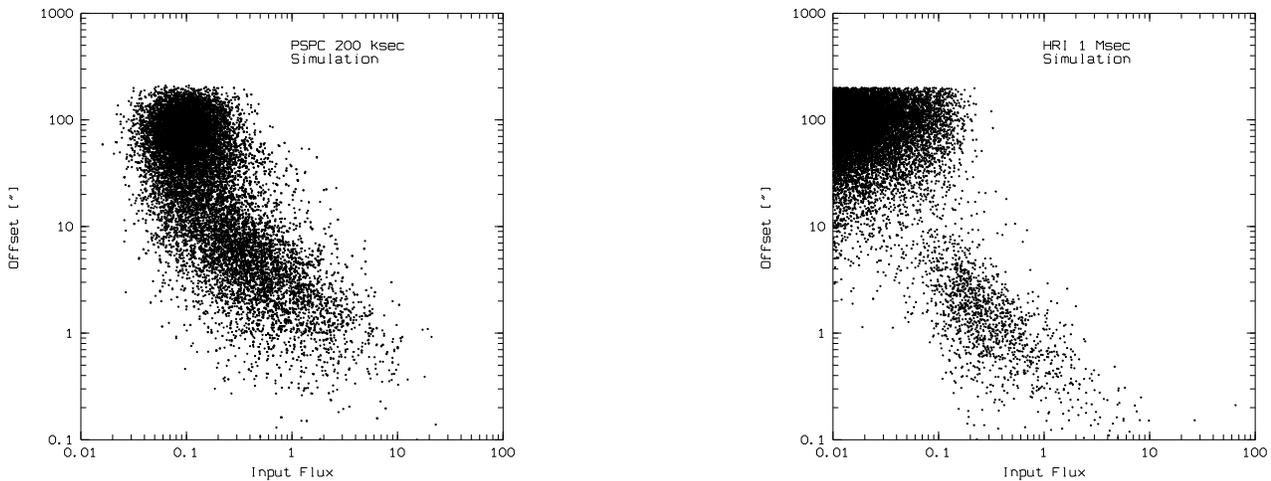

\unitlength1cm
\begin{minipage}[t]{8.0cm}
\begin{picture}(8.0,8.0)
\psfig{figure=h0499.f9,width=8.0cm,angle=-90.}
\end{picture}\par
\end{minipage}
\hfill
\begin{minipage}[t]{8.0cm}
\begin{picture}(8.0,8.0)
\psfig{figure=h0499.f10,width=8.0cm,angle=-90.}
\end{picture}\par
\vskip -0.2 truecm
\end{minipage}
\caption[ ]{\small Position deviation between input source and nearest 
output object as a function of input flux for the simulations in Fig. 
\ref{RNS}. The plot is cut off artificially for distances larger than
200 arcsec.}
\label{ERR}
\end{figure*}

\subsection{Effect of source confusion on optical identification}

While it is obviously possible to correct for confusion effects on the source
counts in a statistical way, the optical identification process relies on the
position of individual sources. 
An independent measure of confusion is the fraction of sources with 
significant positional deviation from their respective counterparts. 
Fig. \ref{ERR} shows for each simulated source the distance to the nearest
output source as a function of simulated input flux. As expected,
bright sources have relatively small positional errors and therefore small
deviations of a few arcsec between simulated and output position. For fainter 
sources there is an incompleteness limit. More and more input 
sources are not detected at all and therefore are matched to the nearest 
(but wrong) detected X--ray source. For the HRI simulation this yields a 
well--defined cloud of spuriously matched sources below a flux of 
$\sim 2 \times 10^{-15}~cgs$. For the PSPC this incompleteness limit sets in
below $\sim 4 \times 10^{-15}~cgs$ and the spurious match cloud is not disjunct
of the detected source correlation track. Confusion leads 
to a halo of PSPC sources with significantly larger positional deviation
than the bulk of detected sources at the same flux. 
A practical limiting radius for the optical identification of faint 
ROSAT X--ray sources is 15", which is used by several authors (see e.g. 
Bower et al., 1996, Georgantopoulos et al., 1996).  
In Table 5 we give the {\it fraction of lost sources} $FL(>15")$ which have no
detected source within a distance of 15" from the input position.

The quantity FL is a good tool to define a reliable flux limit for a survey,
because it combines both the confusion and incompleteness effects. 
For practical applications we need, 
however, a quantity that can be applied to existing surveys, i.e. to 
detected sources. To do this, we invert the question that led to the quantity
FL and ask: how many of the detected sources appear at a detected position
further away from any "reasonable" input position than e.g. 15". For this 
purpose we exclude as reasonable input positions sources with input fluxes 
more than a factor 3 below the detected flux. This gives us an 
estimate of the {\it fraction of unidentifiable sources} in a survey, which we 
denote with the quantity FU. 

The quantities FC, FL and FU are compared in Table 5 for different 
surveys and flux ranges.
This table shows that, as expected, confusion effects increase dramatically 
towards fainter fluxes. For fluxes above the incompleteness limit the  
different confusion measures, FC, FL and FU lead to similar estimates for the 
number of confused sources. Below this limit, FL increases dramatically,
indicating severe incompleteness of the survey in addition to confusion.
Using this table we have chosen the relatively conservative limits in 
flux and off-axis angles employed in the current survey, i.e. a flux 
limit of $5.5 \times 10^{-15}$ for off-axis angles smaller than 12.5 arcmin
and a flux limit of $1.1 \times 10^{-14}$ for off-axis angles between 
12.5 and 18.5 arcmin. In the discussion section we compare these simulations 
to shallower PSPC surveys in the literature.

\section{Discussion}

With a limiting flux of $\sim 10^{-15}~erg~cm^{-2}~s^{-1}$, the
X-ray survey in the {\it Lockman Hole} represents the deepest X-ray survey 
ever performed. The total observing time invested is quite comparable to that
of other major astronomical projects, like e.g. the Hubble Deep Field 
(Williams et al., 1996).
Because of the expected confusion in the PSPC it was clear from the 
beginning, that HRI data would be necessary to augment the PSPC identification
process. Because of the smaller field-of-view of the HRI and because of its 
lower quantum efficiency, it was necessary to invest about a factor five more 
HRI time than PSPC time. Both the HRI raster scan, which provides excellent
positions for all brighter objects in the PSPC field, and the HRI ultradeep
survey in the (slightly offset) central part of the field 
allow almost complete optical identifications of sources down to 
$5.5 \times 10^{-15}~erg~cm^{-2}~s^{-1}$ (see paper II). Apart from this, the 
ultradeep HRI 
data provides a survey in its own right, which is not yet fully exploited. It 
will ultimately lead to reliable optical identifications a factor of 3-4
deeper than the current survey. On the other hand, the Lockman Hole data
are also of fundamental importance for other observations with the ROSAT HRI 
since they provided for the first time a significant determination of the 
HRI scale factor correction which is important for all high-quality 
astrometry with the HRI.

\subsection{The ROSAT log(N)--log(S) function}

The new PSPC and HRI data shown in Fig. \ref{LNS} carry the resolved source 
counts a factor of 2.5 deeper than the most sensitive previous determinations 
(H93, Branduardi-Raymont et al., 1994). It is very 
reassuring to see that the source counts are still consistent with the 
previous fluctuation analyses (H93, Barcons et al., 1994).
There are, however, some problems at the bright end of the source counts.
In our previous paper we used the source counts from the EMSS (Gioia et al., 
1990) to constrain the bright end ($S_{(0.5-2.0 keV)} > 1.8 \times 10^{-13}$)
of the log(N)--log(S) relation. The first 
study to note a discrepancy between the EMSS and ROSAT counts was the RIXOS 
survey. Page et al. (1996) found, that the EMSS log(N)--log(S) for AGN is
significantly lower than the one from RIXOS. The source
counts from the ROSAT All-Sky-Survey Bright Source Catalogue 
(Voges et al., 1996) confirm this trend and extend it to fluxes 
well above $10^{-12}~erg~cm^{-2}~s^{-1}$ (Hasinger et al., 1997, in prep.). 
It appears that the log(N)-log(S) function at fluxes brighter than 
$10^{-13}~erg~cm^{-2}~s^{-1}$ has a slope of $\beta_0 \sim 2.3$ and a 
normalization of $\sim 91$ and is therefore significantly flatter than what 
we assumed in H93  from a fit to the EMSS total source counts. 
The flatter log(N)--log(S) slope at 
bright fluxes has some consequences for the calculation of the resolved 
fraction.

\subsection{The resolved fraction}

As in H93 we restrict the analysis of the resolved fraction to the energy
band 1--2 keV in order to minimize galactic contamination in the X-ray 
background spectrum.
The absolute level and the spectrum of the X-ray background in this energy
range is still a matter of debate (see the discussion in Hasinger 
1996). In H93 we assumed a 1--2 keV background flux of 
$1.25 \times 10^{-8}~erg~cm^{-2}~s^{-1}~sr^{-1}$. 
From ASCA data in the energy range 1-7 keV, Gendreau et al. (1995) 
have derived a flat spectrum with photon index 1.4 and a normalization of 
$8.9~keV~cm^{-2}~s^{-1}~keV^{-1}$ at 1 keV. Integrating this spectrum  
we derive a lower limit to the 1--2 keV XRB flux of 
$1.22 \times 10^{-8}~erg~cm^{-2}~s^{-1}~sr^{-1}$.
This can be contrasted to the earlier ROSAT PSPC determination of a steep 
spectrum in the 0.5--2 keV band, with photon index around 2 and a 
normalization around 13 (see Hasinger 1996).
Integrating this over the 1--2 keV band yields a flux of 
$1.44 \times 10^{-8}~erg~cm^{-2}~s^{-1}~sr^{-1}$, which is consistent with
the background flux derived by Chen et al., 1997 from a joint ROSAT/ASCA
fit (1.46 in the same units).

In H93 we integrated the analytic log(N)--log(S) function
(Eq. 3) for fluxes brighter than $2.5 \times 10^{-15}~erg~cm^{-2}~s^{-1}$,
including a fit to the EMSS at fluxes brighter than 
$1.8 \times 10^{-13}~erg~cm^{-2}~s^{-1}$. We obtained a resolved 
1--2 keV flux of $0.74 \times 10^{-8}~erg~cm^{-2}~s^{-1}~sr^{-1}$.
If we follow the same prescription, but now integrating the H93 source counts 
above our limiting flux of $10^{-15}~erg~cm^{-2}~s^{-1}$, we arrive at
a resolved flux of 0.89 in the same units. If we, however, correct for the 
flatter slope of the bright end log(N)--log(S) (see above), we arrive at a 
resolved flux value of 0.99 (same units). We use this resolved flux in
comparing with the total diffuse background in the 1--2 keV band.

\begin{table*}
\caption[ ]{Comparison of ROSAT surveys}
\begin{center}
\begin{tabular}{lccclrcrrr}
\noalign{\smallskip}
\hline
\noalign{\smallskip}
Survey$^{\rm a}$ & Area$^{\rm b}$ & T$^{\rm c}$
& $S_{lim}^{\rm d}$ & Off$^{\rm e}$ & Err$^{\rm f}$ &
$S_{range}^{\rm g}$ & FC$^{\rm h}$ & FL$^{\rm i}$ & FU$^{\rm j}$\\
\noalign{\smallskip}
\hline
\noalign{\smallskip}
 RIXOS  &$\sim20$& $>$8    & 3.0  \\
 CRSS   & 3.9 & $>$6       & 2.0  \\
 DRS    & 1.4 &  21-49     & 0.3 & 15   & 15 & 0.3-0.6 & 13 & 62 & 23\\
 MARA   & 0.2 &  55        & 0.5 & 15   & 15 & 0.5-1   & 11 & 33 & 19\\
 NEP    & 0.2 &  79        & 1.0 & 15.5 & 15 & 1-2     &  6 &  6 &  7\\
 UKDS   & 0.2 & 110        & 0.2 & 15   & 10 & 0.2-0.4 & 10 & 76 & 42\\
 RDS    & 0.3 & 207        & 0.5 & 12.5 & 15 & 0.5-1   &  7 & 10 &  5\\
 HRI    & 0.1 &1000        & 0.2 & 12.5 &  5 & 0.2-0.4 &  2 &  7 &  5\\
\noalign{\smallskip}
\hline
\end{tabular}
\end{center}
\begin{list}{}{}
{\small
\item[$^{\rm a}$] cf. text for explanation of acronyms
\item[$^{\rm b}$] approximate, in square degrees
\item[$^{\rm c}$] PSPC exposure time, in ksec
\item[$^{\rm d}$] quoted limiting flux $S(0.5-2\rm{keV})$ in units
 of $10^{-14}$ erg cm$^{-2}$ s$^{-1}$
\item[$^{\rm e}$] maximum off-axis angle in survey in arcmin 
\item[$^{\rm f}$] radius of error circle searched in survey in arcsec 
\item[$^{\rm g}$] flux range for which FC, FL and FU  have been estimated 
\item[$^{\rm h}$] percentage FC of contaminated objects 
\item[$^{\rm i}$] percentage FL of lost objects 
\item[$^{\rm j}$] percentage FU of unidentifiable objects 
}
\end{list} 
\end{table*}

If we adopt a 1--2 keV background flux of
$1.45 \times 10^{-8}~erg~cm^{-2}~s^{-1}~sr^{-1}$, we have resolved 68\%
of the 1--2 keV X--ray background at a flux of $10^{-15}~erg~cm^{-2}~s^{-1}$.
If, however, the lower ASCA spectrum with a flux of 
$1.22 \times 10^{-8}~erg~cm^{-2}~s^{-1}~sr^{-1}$ holds,
we have already resolved 81\% of the background.  We see, that the uncertainty
in the resolved fraction is now dominated by the systematic error in the 
absolute background flux and not by the source counts. 
As a best guess for the resolved fraction we take $70-80\%$.

\subsection{Comparison to shallower PSPC surveys}

The optical identifications in the {\it Lockman Hole} and the detailed
comparison with other work is described in paper II
(Schmidt et al., 1997). Here we want to draw global comparisons
with other PSPC surveys, in particular applying the 
simulation results described above.  
Quite a number of groups are involved in optical identifications of 
deep and medium-deep survey fields observed with ROSAT (see Table
6) and already many 
papers have been written about the interpretation of ROSAT survey 
results. Among the most debated findings
is the claimed detection of a possible new class of X--ray active,
optically relatively normal emission line galaxies at faint flux
levels (Griffiths et al., 1996, McHardy et al., 1997).  
Unfortunately, however, only very few surveys so far have been published 
formally (i.e. including a catalogue of detected sources and a 
detailed description of the detection and identification procedure),
so that quantitative comparisons can be made. Among those published
are the Cambridge-Cambridge ROSAT Serendipity Survey (CRSS, Boyle et 
al., 1995), a small part of the deep ROSAT survey (DRS, Shanks et al., 1991,
see also Georgantopoulos et al., 1996) and the North Ecliptic Pole Survey 
(NEP, Bower et al., 1996). Among those waiting to be fully published 
are the ROSAT International X--ray Optical Survey (RIXOS, Mason et al., 
in prep., see also Page et al., 1996), the UK ROSAT deep field survey
(UKDS, McHardy et al., 1997), the full DRS,  the survey in the Marano field 
(MARA, Zamorani et al.,
in prep.), and finally our own ROSAT Deep Survey (RDS, this paper;
Schmidt et al., 1997). Nevertheless, the global properties of these surveys
are known and some details can be obtained from the existing literature.
See Table 6 for a summary of the surveys in question. The table also 
includes a prediction for the HRI ultradeep survey. 

Using the experience gained from the ROSAT Deep Survey and applying the 
simulation tools to the shallower surveys we can now make some statements
about the expected quality of the other surveys, in particular with 
respect to possible confusion and the corresponding optical
misidentification. As we have seen in Table 5, various confusion 
problems, as indicated by the FL, FC and FU percentages, become severe roughly 
within a factor of 2 from the formal detection 
threshold of a PSPC exposure longer than 50 ksec. It is interesting to note 
that this behaviour
is not a strong function of the exposure time. In order to obtain a 
more quantitative assessment of the different surveys, we use our PSPC
simulations of 50 ksec (DRS, MARA), 110 ksec (UKDS) and 200 ksec (RDS)
exposures applying as far as possible the detailed information about exposure 
times, quoted flux limit, off-axis angles and assumed error circle radii in 
the individual surveys (see Table 6). For the 80 ksec NEP survey we use 
quantities interpolated between the 50 ksec and the 110 ksec simulations.  
For lack of even shallower simulations we do not make statements about 
the RIXOS and CRSS surveys.
For each of the deep and medium-deep surveys in Table 6 we calculate the 
quantities FC, FL and FU, for a flux range within a 
factor of two from the sensitivity limit quoted by the authors. This way 
we can predict for each survey the fraction of misidentified sources among 
the faint source population, where presumably the new discoveries would
be expected.

It is not surprising to see, that those surveys that employ a relatively 
conservative limiting sensitivity (as judged from the ratio between 
exposure time and flux limit) have very small fractions of unidentifiable 
sources. This is confirmed by the high rate of identifications e.g
for the NEP survey (Bower et al., 1996) and the RDS (Schmidt et al., 1997).
The RDS is additionally helped by the HRI data in the field.
Larger fractions of unidentifiable sources on the order of 10\% are expected
for the DRS and the Marano survey. Indeed, in the Marano field, which 
unfortunately does not have HRI coverage, we see a significant number of 
empty error boxes (Zamorani et al., in preparation). Unidentified source
fractions of order 20\% are no major problem as long as one is studying 
majority populations of X--ray sources (e.g. AGN). They are, however, 
already a substantial problem if one tries to identify new classes of
objects which necessarily are minority classes. There the error can 
easily approach 100\%. For the UKDS, which tried to push
deepest in the optical identification, unfortunately the largest 
unidentifiable fraction (42\%) is
expected. Indeed, about 26\% out of a total of 34 UKDS sources fainter than
$5 \times 10^{-15}~erg~cm^{-2}~s^{-1}$ are unidentified (as derived from 
Fig. 5 in McHardy et al., 1997), either because the error boxes 
are empty or contain too faint objects. Judging from our simulations
we also expect some of the UKDS sources with proposed optical counterparts
to be misidentified. 

\section{Conclusions}
 
We have presented the complete catalogue of 50 X--ray sources with 0.5--2 keV  
fluxes above $ 5.5 \times 10^{-15}~erg~cm^{-2}~s^{-1}$ from the ROSAT Deep Survey 
pointing of exposure time 207 ksec in the Lockman Hole. The X--ray positions  are
largely defined by additional ROSAT HRI observations of more than 1Msec total 
exposure time. Using the HRI data we have derived a new log(N)--log(S) function reaching 
a source density of $970 \pm 150 ~deg^{-2}$ 
at a limiting flux of about $10^{-15}~erg~cm^{-2}~s^{-1}$. 
At this level 70-80\% 
of the 0.5--2 keV X--ray background is resolved into discrete sources. 
The observations and the analysis procedure specifically developed to 
cope with confused X--ray observations have greatly profited from detailed 
simulations of PSPC and HRI fields. Based on these simulations 
we have defined conservative limits on flux and on off-axis angles, which 
guarantee a high reliability of the catalogue. The fraction of confused or 
unidentifiable sources in the catalogue is expected at only a few percent, 
but is probably even lower due to the existence of the deep HRI data. 

We also discussed simulations of shallower fields and show that surveys, 
which are only based on PSPC exposures larger than 50 ksec, become
severely confusion limited typically a factor of 2 above their $4\sigma$ 
detection threshold. Sizeable fractions of confused and unidentifiable 
sources are expected for some of the published surveys. This may have
consequences 
for some recent claims of a possible new source population emerging at the 
faintest X--ray fluxes.

\begin{acknowledgements}
The ROSAT project is supported by the Bundesministerium
f\"ur Forschung und Technologie (BMFT), by the National
Aeronautics and Space Administration (NASA), and the Science
and Engineering Research Council (SERC). This work was supported
in part by NASA grants NAG5--1531 (M.S.), NAG8--794, NAG5--1649,
and NAGW--2508 (R.B. and R.G.). G.H. acknowledges the grant 
FKZ 50 OR 9403 5 by the Deutsche Agentur f\"ur Raumfahrtangelegenheiten
(DARA). G.Z. acknowledges partial support by the Italian 
Space Agency (ASI) under contract ASI 95--RS--152.

\end{acknowledgements}

\end{document}